\documentclass[preprint,showpacs,preprintnumbers,amsmath,amssymb,aps,prb]{revtex4}
\usepackage{graphicx}
\usepackage{times}
\usepackage{subfigure}
\usepackage{epsfig}
\begin{document}
\preprint{LPA-001}


\title{Hole-LO phonon interaction in InAs/GaAs quantum dots\\}

\author{V.\ Preisler}
\author{R.\ Ferreira}
\author{S.\ Hameau}
\author{L.A.\ de\ Vaulchier}
\author{Y.\ Guldner}

\affiliation{Laboratoire Pierre Aigrain, Ecole Normale
Sup\'erieure, 24 rue Lhomond, 75231 Paris Cedex 05, France}

\author{M.\ Sadowski}

\affiliation{Grenoble High Magnetic Field Laboratory, CNRS/MPI, 25
avenue des Martyrs, 38042 Grenoble Cedex  9, France}

\author{A.\ Lemaitre}

\affiliation{Laboratoire de Photonique et Nanostructures, Route de
Nozay, 91460 Marcoussis, France\\}


\begin{abstract}
We investigate the valence intraband transitions in p-doped
self-assembled InAs quantum dots using far-infrared
magneto-optical technique with polarized radiation.  We show that
a purely electronic model is unable to account for the
experimental data.  We calculate the coupling between the confined
hole and the longitudinal optical phonons of the surrounding
lattice using the Fr\"ohlich Hamiltonian, from which we determine
the polaron states as well as the energies and oscillator
strengths of the valence intraband transitions. The good fit
between the experiments and calculations provides strong evidence
for the existence of hole-polarons and demonstrates that the
intraband magneto-optical transitions occur between polaron
states.
\end{abstract}


\date{\today}
\maketitle


\section{introduction}
Carrier-phonon interactions in quantum dots (QDs) have attracted
considerable attention recently because they are essential in
understanding the electronic properties of such systems, for
instance the carrier relaxation which is of particular interest in
QDs.  Various experimental and theoretical results have
demonstrated that carriers confined in InAs/GaAs QDs are strongly
coupled to the longitudinal optical (LO) vibrations of the
underlying semiconductor
lattice\cite{hameau02,sarkar05,knipp97,hameau99,verzelen00,inoshita97,li98}.
For conduction electrons, this coupling leads to the formation of
the so-called electron polarons which are the true excitations of
a charged dot. Far-infrared (FIR) absorption probes directly the
polaron levels instead of the purely electronic states and the
electron polarons have been extensively studied by intraband
magneto-optical transitions in n-doped
QDs\cite{hameau02,hameau99}. But to date, there were few
investigations of valence intraband magneto-optical transitions in
QDs and therefore no direct evidence for the formation of hole
magneto-polarons.

	In this present work, we have studied the hole excitation of
p-doped self-assembled InAs/GaAs QDs using FIR magnetotransmission
experiments up to 28 tesla at 2 K.  We have investigated the
intraband transitions between the ground and the first excited
valence states.  Depending on the FIR polarization, two different
transitions can be excited, whose intensity and dispersion versus
magnetic field show strong deviation with respect to the
predictions of a purely electronic level model. Using the
Fr\"ohlich Hamiltonian, we have calculated the coupling between
the low-lying confined hole states, which we find using a simple
one-band model, and the lattice modes.  We have determined the
hole polaron states as well as the energies and oscillator
strengths of the intraband transitions. We show in this work that
our model fits the experimental data very well, demonstrating that
the measured transitions occur between hole polaron states.

\section{Experimental Details}
The InAs/GaAs dots investigated here were grown on (001) GaAs
substrates by molecular-beam epitaxy using the Stranksi-Krastanov
growth mode of InAs on GaAs\cite{Goldstein85}.  As the resonant
FIR absorption associated with a single dot layer is weak (a few
0.1\%)\cite{hameau02,fricke96}, samples containing a multistack of
20 layers of InAs QDs were prepared in order to strengthen this
absorption. Each dot layer is separated by a 50-nm-thick GaAs
barrier.  The density of QDs is $\sim 4\times
10^{10}\,\text{cm}^{-2}$ for each plane, corresponding to an
average center to center distance of 50 nm, large enough to
neglect any interdot interactions.  We can, therefore, reasonably
consider our samples as consisting of isolated QDs. The gaps of
the QDs were measured in the different samples by
photoluminescence (PL) experiments at 4 K. The PL peak is centered
at $\sim1.2\text{ meV}$, which is a typical value for lenslike
InAs islands with a height of $\sim\text{2-3 nm}$ and a lateral
diameter of $\sim\text{20 nm}$. The dot filling is realized by a
Be delta doping of each GaAs barrier at 2 nm under each dot layer.
In the following we discuss results obtained from two samples
labelled (I) and (II). Sample (I) has a doping level of $\sim
5\times 10^{10}\,\text{cm}^{-2}$, whereas sample (II) has a doping
level of $\sim10\times 10^{10}\,\text{cm}^{-2}$. The doping level
was adjusted to transfer on average one or two holes per dot and
populate only the ground state.  We have investigated the optical
transitions between the ground state and first excited states.
Such transitions, which involve only valence-band states,
correspond to resonance energy in the FIR.  The sample
transmission at 2 K was recorded by Fourier-transform spectroscopy
in the FIR range ($50-700\,\textrm{ cm}^{-1}$). In order to
eliminate optical setup effects, the sample transmission was
normalized to the substrate transmission.  Two types of
magnetotransmission experiments have been performed with the
radiation propagating perpendicular to the QD layers of the
samples: measurements up to B=15 T using a superconducting magnet
with linearly polarized light and measurements up to B=28 T at the
Grenoble High Magnetic Field Laboratory using a resistive magnet
and non-polarized light. The magnetic field was applied parallel
to the growth axis (Faraday configuration).
\begin{figure}[b]
\centering
\subfigure{\includegraphics[width=0.2161\textwidth]{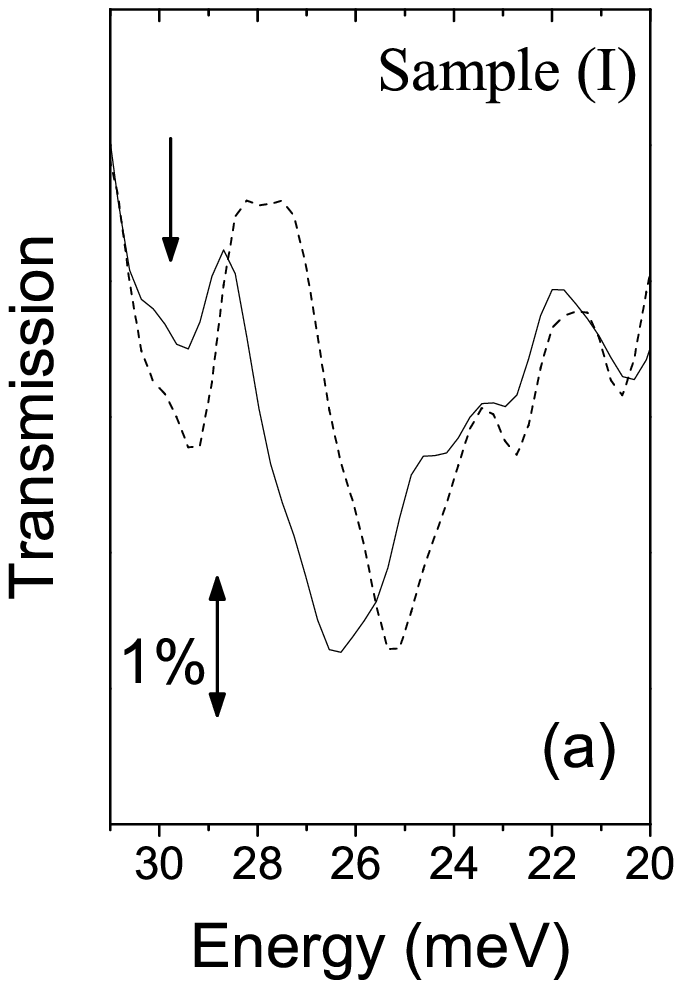}
\label{fig1a}}
\subfigure{\includegraphics[width=0.225\textwidth]{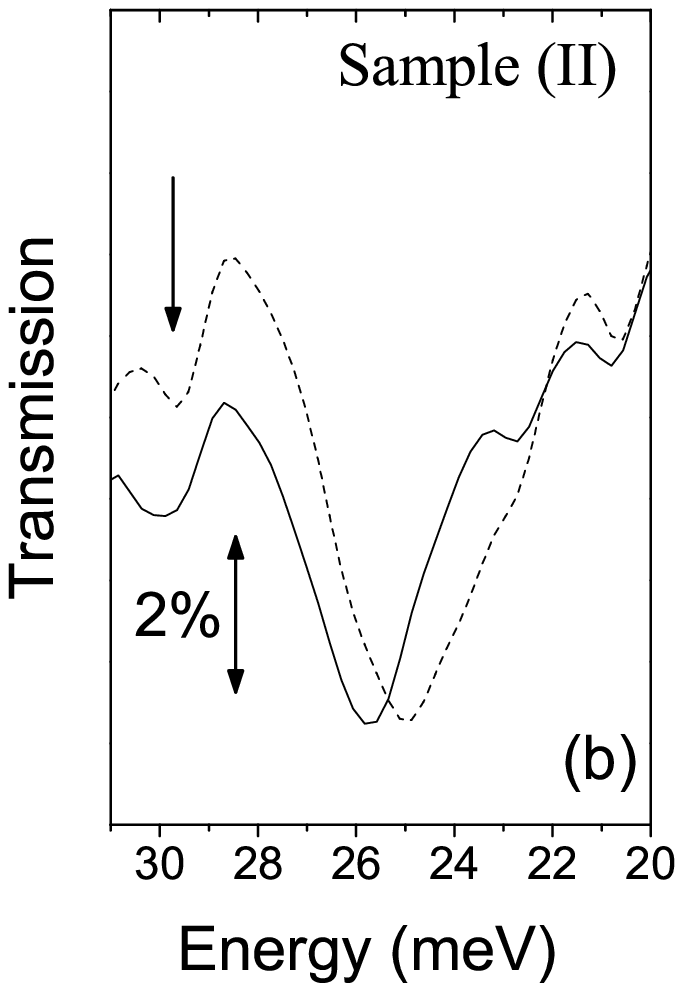}
\label{fig1b}}
 \caption{Transmission spectra at B=0 for radiation
linearly polarized along the [110] (solid curve) and the
[1$\overline1$0] (dashed curve) directions for sample (I) (a) and
sample (II) (b). The arrows indicate the absorption associated
with the LO-phonon of InAs. } \label{fig1}
\end{figure}
\section{Results}
\begin{figure}[t]
 \subfigure{\includegraphics[width=0.3\textwidth]{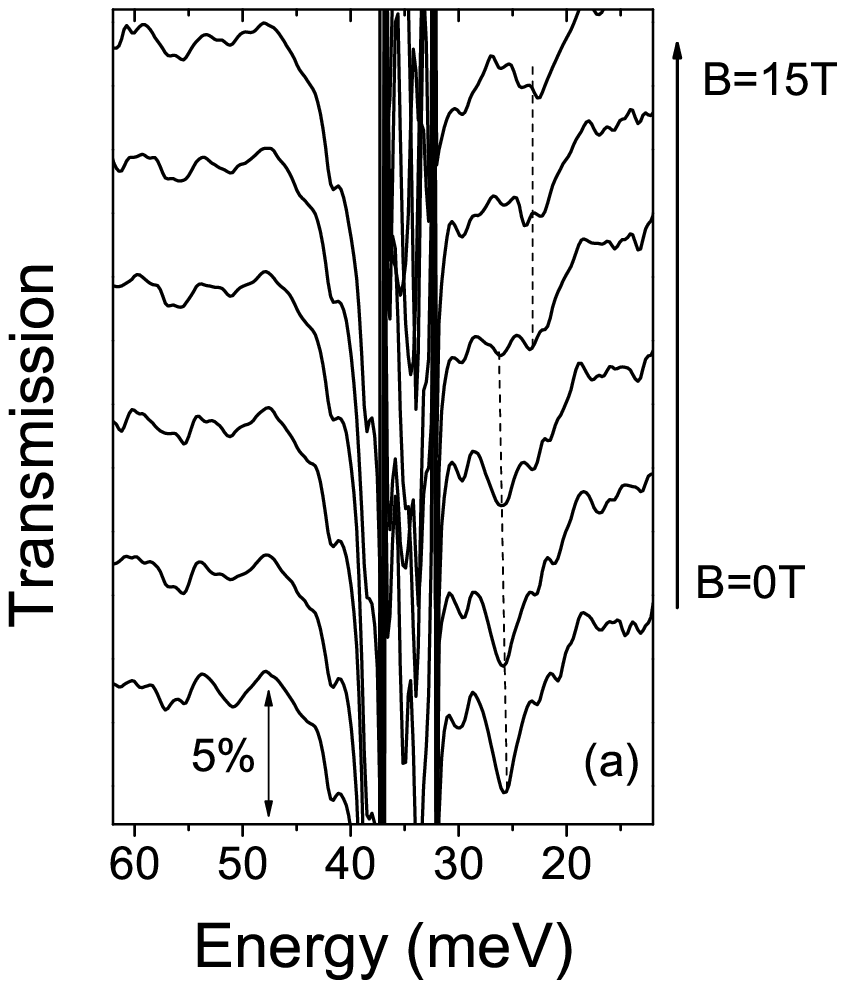}
  \label{fig2a}}
\subfigure{\includegraphics[width=0.3\textwidth]{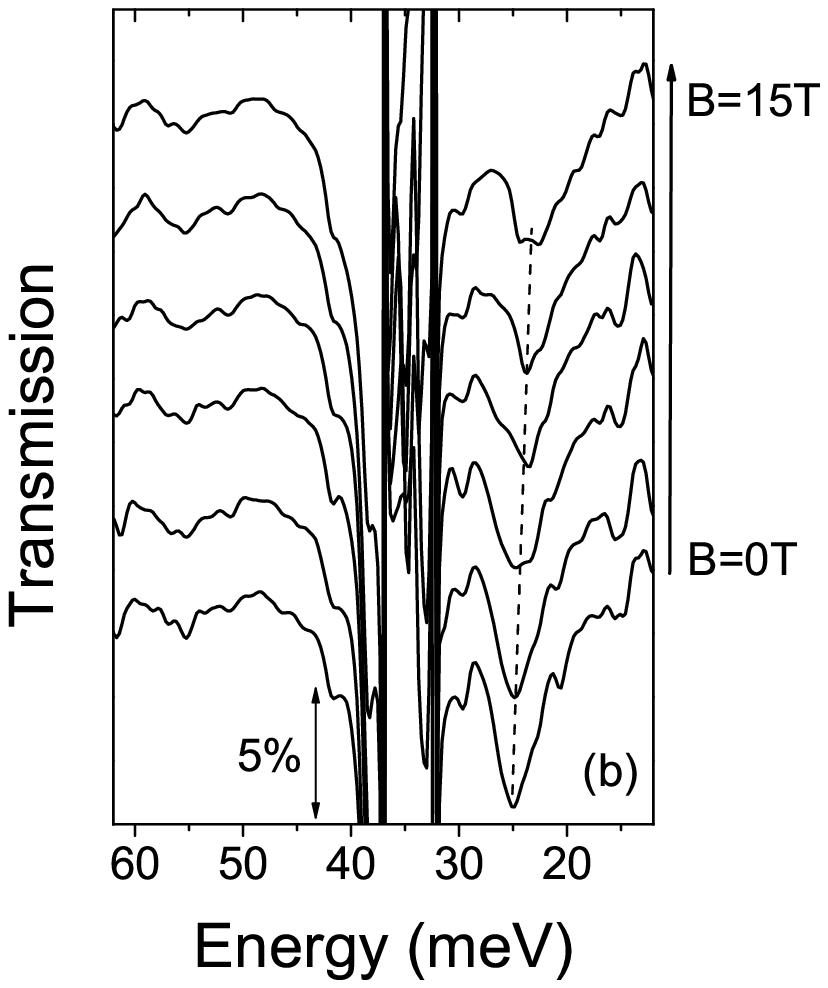}
\label{fig2b}}
 \caption{Magnetotransmission spectra measured in sample (II) for
radiation linearly polarized along the (a) [110] and (b)
[1$\overline1$0] directions and recorded at 2 K from B=0 to B=15 T
every 3 T. Traces have been vertically offset for
clarity.}\label{fig2}
\end{figure}

Intraband transitions in InAs/GaAs self-assembled QDs usually
display anisotropic behavior when the FIR radiation is polarized
along either the [110] or the [1$\overline1$0]
direction\cite{hameau02}. Such behavior could arise, e.g., from an
anisotropy of the dot shape\cite{fricke96,hasegawa98}, interband
and piezoelectric coupling\cite{stier01}, or atomistic
contributions\cite{bester03}.  We have therefore performed
magnetotransmission spectra for FIR radiation, linearly polarized
along the [110] and [1$\overline1$0] directions of the sample.
Fig.~\ref{fig1} displays the FIR absorption spectra at zero
magnetic field recorded for the two polarizations in sample (I)
(Fig.~\ref{fig1a}) and sample (II) (Fig.~\ref{fig1b}).  A main
absorption is observed at an energy of $\sim$ 26 meV in both
samples for the [110] polarization.  For the [1$\overline1$0]
polarization, the main absorption occurs at a slightly lower
energy.  The anisotropy related energy splitting is found to be
$\sim$ 1.2 meV in sample (I) and $\sim$ 0.8 meV in sample (II).
Such values are significantly less than those measured for
intraband transitions in n-type samples grown in similar
conditions.  The sharpness of the lines [the full width at half
maximum (FWHM) is $\sim\text{3 meV}$] is good evidence of the high
quality of these samples.  The main line intensity is $\sim$ 2$\%$
for sample (I) and $\sim$ 4$\%$ for sample (II).  Such a
difference in intensity can be explained by the different doping
levels of the two samples, as described for instance in Appendix A
of reference \onlinecite{hameau02}. Note, as well, that a smaller
absorption is observed in both samples at around 29 meV (indicated
by the arrows in Fig.~\ref{fig1}).  We have verified that this
absorption is also measured in undoped samples grown in similar
conditions, while the other absorptions associated with the doped
dots are not observed. We thus associate this $\sim 1\%$ sharp
absorption with the InAs like LO phonon. Fig.~\ref{fig2a} displays
the magnetotransmission spectra for radiation along the [110]
direction of sample (II) and recorded at 2 K from B=0 to B=15 T
every 3 T. As the magnetic field increases, the main absorption
observed at zero magnetic field at $\sim$ 26 meV stays nearly
constant in energy all the while decreasing in intensity. At
around 6 tesla, a second lower energy absorption appears at 23
meV. This second absorption decreases in energy as the magnetic
field is increased. The zero transmission region from 32 meV to 37
meV corresponds to the restrahlen band of the GaAs substrate.

Let us now compare this behavior with the magnetotransmission
spectra for radiation polarized along the [1$\overline1$0]
direction. Fig.~\ref{fig2b} displays the magnetotransmission
spectra for radiation polarized along the [1$\overline1$0]
direction with an applied magnetic field. The absorption minimum
decreases in energy as the magnetic field increases while its
oscillator strength stays strong for all magnetic fields.

\begin{figure}[t]
\includegraphics[width=0.5\textwidth]{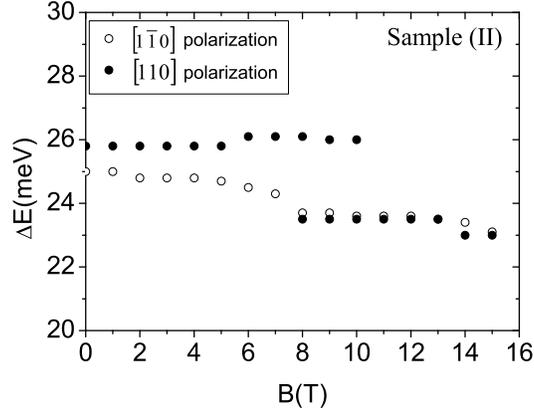}
\caption{Magnetic field dispersion of the resonances in sample
(II) for the two polarization directions; [110] in full circles
and [1$\overline1$0] in open circles.} \label{fig3}
\end{figure}

The different behavior between these two polarizations is clearly
demonstrated in Fig.~\ref{fig3} where we have plotted the energy
absorption minima of the two polarizations as a function of
magnetic field. We see that one polarization contains one unique
absorption that decreases in energy while the other polarization
has two absorptions: a high energy branch observable at low
magnetic field and low energy branch observable at high magnetic
fields. It is clear that the transition energies we observe
experimentally cannot be accounted for by a simple Zeeman
splitting effect that we would expect in a purely electronic
model.  Indeed, the lower and upper branches would be symmetric in
that case with similar oscillator strengths.  We observe quite
similar magnetotransmission results for sample (I).

\section{Analysis and Discussion}
In polar semiconductors, such as III-V materials, anomalies
observed for intraband transitions often originate in carrier-LO
phonon interactions when the energy separation between the ground
and excited states approaches the LO phonon energy
$\hbar\omega_{LO}$.  The energy of LO phonons in bulk InAs is
$\hbar\omega_{LO}=29\text{ meV}$ which is very close to that of
the upper branch (Fig.~\ref{fig3}).  As mentioned above, a
significant absorption, that is independent of the magnetic field,
is observed around 29 meV in the FIR spectra of all our samples.
Because a purely electronic level model is unable to explain the
experimental data, we have to consider FIR magneto-optical
transitions between polaron states and, therefore, to calculate
the coupling between the relevant mixed hole-phonon states.  In
bulk materials, it is well-known that one has to diagonalize the
Fr\"ohlich Hamiltonian that describes the Coulomb interaction
between a moving charge and the dipoles vibrating at the angular
frequency $\omega_{LO}$ corresponding to a longitudinal optical
mode.  This mode is associated with the partial ionicity of the
bonds between the two different atoms that constitute the III-V
semiconductors.  In QDs, the phonon spectra are not known
accurately as a result of the uncertainties of the shape and
composition of the actual dots. But since an actual InAs island
consists of several thousand unit cells we can expect the dot to
have a quasibulk phonon spectra. In addition, like in III-V bulk,
each anion is surrounded by four cations with a slightly polar
bond between them.  Therefore, the basic ingredients of the
Fr\"ohlich Hamiltonian are maintained in actual dots. These
considerations, along with the fact that it works very well for
n-doped samples\cite{hameau02}, have led us to take a bulklike
Fr\"ohlich Hamiltonian to describe the interaction between a hole
bound to a dot and the optical phonon mode of the structure.

\begin{figure} \subfigure{
\includegraphics[width=0.5\textwidth]{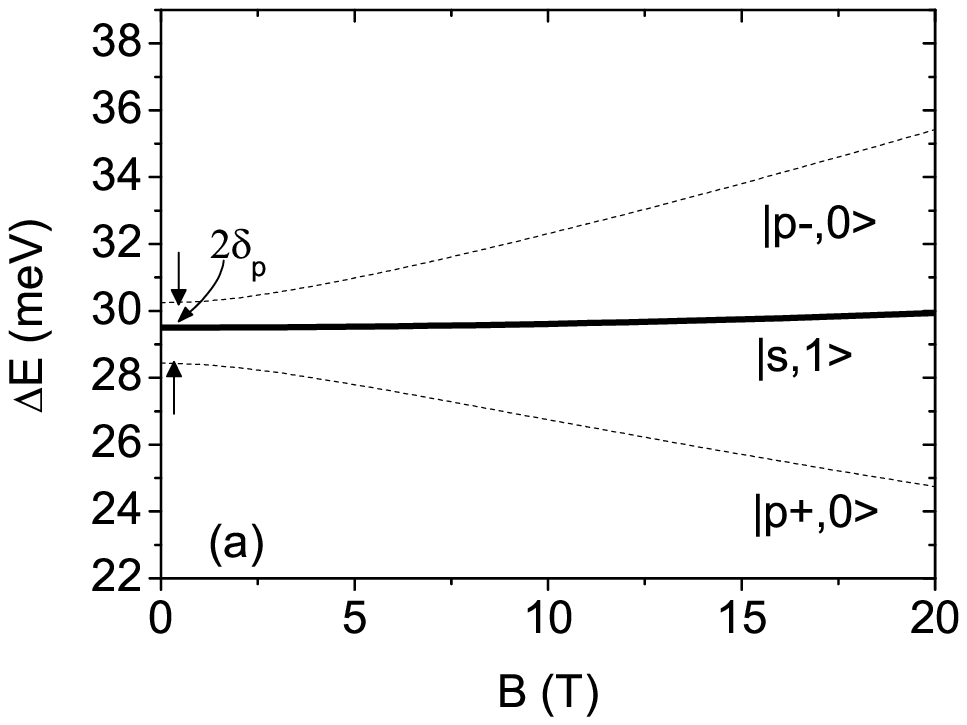}
 \label{fig4a}}

\subfigure{
\includegraphics[width=0.5\textwidth]{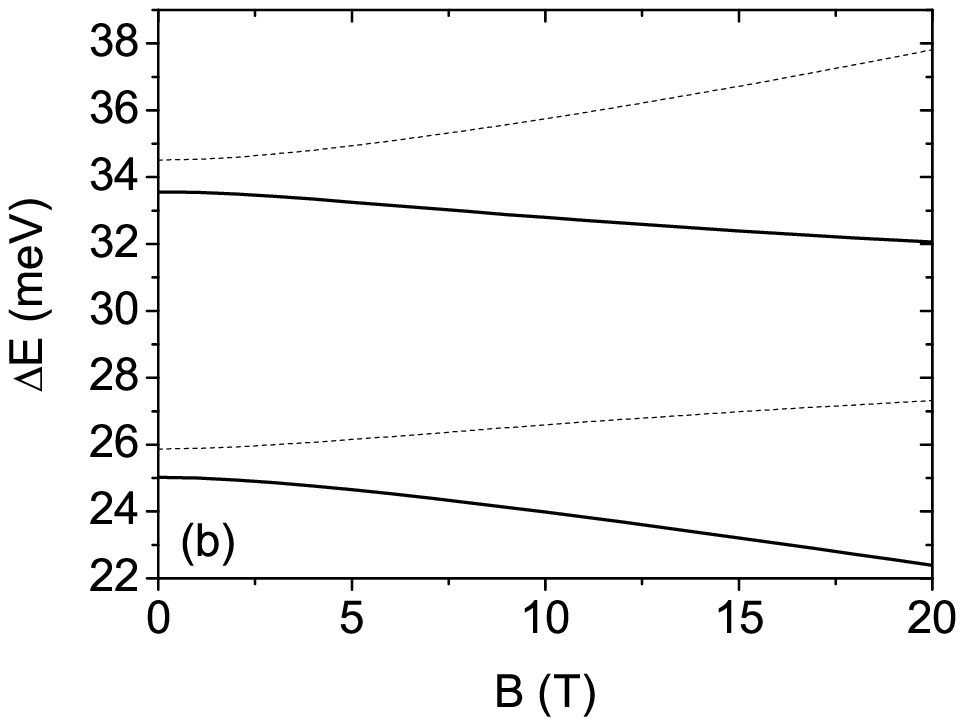}
 \label{fig4b}}
 \caption{Resonance dispersions calculated from the numerical
 diagonalization of the Hamiltonian without coupling term (a) and
 with Fr\"ohlich term (b) for the parameters listed in the text.}
\end{figure}
	We evaluate the hole states within a one-band (parabolic) model.
Such a model gives an accurate description of the coupling between
electrons and phonon modes in n-doped samples\cite{hameau02}.
However, the use of such a simple model is less evident for the
valence levels of semiconductor nanostructures, as discussed in
many works (see e.g. references \onlinecite{stier01,bester03} and
references there in).  For instance, non-parabolicity and mass
anisotropy should play an important role in the description of the
hole states.  In order to account for these effects in a
simplified way, we consider an anisotropic dispersion for heavy
holes, with an in-plane mass chosen to best fit the
magneto-optical experimental results.  Of course, a more rigorous
theory should incorporate such mass effects from the onset.
However, as shown below, our model suffices to provide a good
description of our experimental results using reasonable fitting
parameters.

	For QDs with a perfect cylindrical symmetry the ground and first
excited states are $s$-like and $p$-like, respectively.  By using
a variational procedure with Gaussian functions\cite{ferreira99},
we have calculated the energy levels of holes in a QD modelled by
a truncated cone of height $h$ and with a circular basis of radius
$R$ and basis angle $30^\circ$ as estimated from transmission
electron micrographs.  We have considered a homogeneous gallium
content of 30\% in the QDs, resulting in a valence band offset of
217 meV\cite{hameau02} and added an anisotropy term in order to
account for the zero field splitting we observe experimentally.
The noninteracting mixed hole-phonon states in an uncoupled system
are labelled $|\upsilon,n_{\textbf{q}}\rangle$ where
$|\upsilon\rangle=|s\rangle,|p_{\pm}\rangle$ are purely electronic
levels (in what follows $p_{\pm}$ denotes the two electronic
levels that result from the two excited states admixed by the
anisotropy term). $|n_{\textbf{{q}}}\rangle$ denotes the ensemble
of the $n$ LO-phonon states in the $\{\textbf{q\}}$ modes.   In
Fig.~\ref{fig4a} we present the calculated energy transitions for
the uncoupled system, where the zero energy has been taken at the
ground state ($|s,0\rangle$) energy.  The splitting,
$2\delta_{p}$, between the two $p$-like components at zero
magnetic field arises from the anisotropy term. The parameters
used in this calculation are described below.

	Now we add the Fr\"ohlich Hamiltonian term which couples states
that differ by one phonon.  As discussed in Appendix A, the
numerical diagonalization of the Hamiltonian including the
Fr\"ohlich term gives the polaron states.  We have taken the
dimensionless Fr\"ohlich constant $\alpha$ that characterizes the
ionicity of the material and the intensity of the coupling as an
adjustable parameter.  It has previously been reported that
$\alpha$ could be enhanced in dots as compared to the bulk
situation\cite{Heitz97}. Fig.~\ref{fig4b} displays the calculated
polaron levels.  We observe four energy transitions: two
transitions represented by dashed lines that are mainly a result
of an interaction between the $|s,1\rangle$ continuum and the
$|p-,0\rangle$ state and two transitions represented by solid
lines that are mainly the result of an interaction between the
$|s,1\rangle$ continuum and $|p+,0\rangle$ state (see Appendix A
for more details).

\begin{figure}[t]
\includegraphics[width=0.5\textwidth]{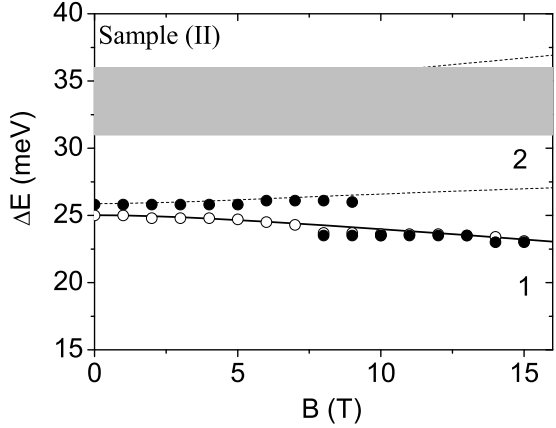}

 \caption{Magnetic field dispersion of the resonances observed for sample(II)
  for the two polarization directions; [110] in full circles and
[1$\overline1$0] in open circles, with the calculated energy
transitions in bold and dashed curves.  The grey area between 32
meV and 36 meV represented the zero transmission region of the
substrate.}
 \label{fig5}
\end{figure}

	Let us now compare our calculated polaron levels with our
experimental data.  The solid and dashed curves of Fig.~\ref{fig5}
are the resonance dispersion calculated for
$\alpha=0.13,m^{*}=0.22 m_{0}\text{ and the anisotropy term
}\delta_{p}=0.9\text{ meV}.$  A dot with a height $h\text{=25.4
\AA}$ and radius $R\text{=91\AA}$ was used for this fit (Note that
these parameters are also used in the calculations shown in
Fig.~\ref{fig4a} and Fig.~\ref{fig4b}). A similar value of the
Fr\"ohlich constant, $\alpha$, has been determined in n-doped
QDs\cite{hameau99}. The energy positions versus field are very
well described by our model. The two higher energy polarons
branches, predicted in Fig.~\ref{fig4b}, are not observed
experimentally in this field range because their energies coincide
with that of the restrahlen band of the GaAs substrate, (grey area
in Fig.~\ref{fig5}).

	In order to explain the variation in intensity we observe
experimentally, we have calculated the oscillator strengths of the
optical transitions between polaron levels as a function of
magnetic field (Fig.~\ref{fig6}).  The equations used in these
calculations are given at the end of Appendix A. The solid curves
represent the evolution of the oscillator strength of the lower
energy branch (labelled (1) in Fig.~\ref{fig5}) whereas the dashed
curves represents the evolution of the higher energy branch
(labelled (2) in Fig.~\ref{fig5}). In addition, the full circles
correspond to light polarized along the [110] direction and the
open circles to a polarization along the [1$\overline1$0]
direction.  Note that, at zero magnetic field, the $\sim\text4\%$
absorption measured in sample (II) for light polarized along the
[1$\overline{1}0$] direction (Fig.~\ref{fig1b}), corresponds to an
oscillator strength of $\sim\text0.6$.  This is corroborated by
the $\sim\text3\%$ absorption for light polarized along the [110]
direction which corresponds to an oscillator strength of $\sim
\text0.45$. Taking into account the signal to noise ratio, which
allows the detection of $\sim\text1\%$ transmission variation, our
experimental sensitivity corresponds roughly to an oscillator
strength of $\sim\text{0.15}$. At zero tesla, our model predicts
the existence of the higher energy branch for the [110]
polarization and the lower energy branch for the [1$\overline1$0]
polarization. As the magnetic field is increased, the oscillator
strength of the high energy branch decreases towards the
experimental limit 0.15.  On the contrary, the oscillator strength
of the lower energy branch increases above the experimentally
observable intensity around 6 T. For the other polarization
[1$\overline1$0], our model predicts one unique observable
absorption with an oscillator strength that stays nearly constant
with the changing magnetic field. This is a good description of
the oscillator strength behavior we observe in the
magnetotransmission spectra of our samples. In addition, we note
that a purely electronic model cannot predict the variation in
oscillator strength with the magnetic field that we observe
experimentally.  The good agreement obtained for the energy
transitions as well as for the evolution of oscillator strengths
demonstrates that the magneto-optical transitions occur between
polaron states.
 \begin{figure}[t]
\includegraphics[width=0.5\textwidth]{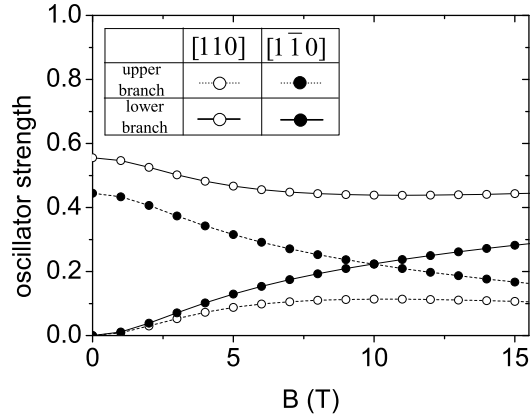}

 \caption{Calculated magnetic field dependence of the oscillator strength for
  the high energy polaron (dashed line) and the low energy polaron
  (solid line) for light polarized along the [110] (full circles)
  and [1$\overline1$0](open circles) directions.}
   \label{fig6}
\end{figure}
\begin{figure}[t]
\includegraphics[width=0.5\textwidth]{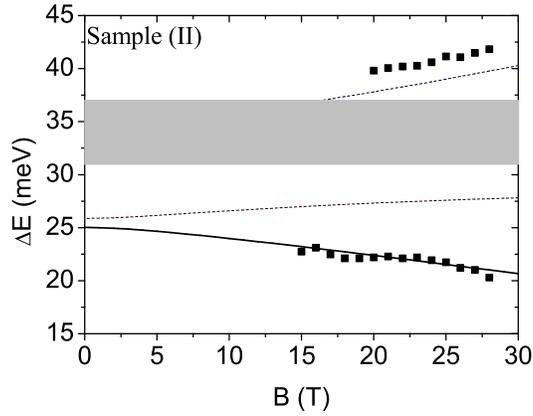}
\caption{High magnetic-field dispersion of the resonances (full
squares) in unpolarized light for sample (II).  The lines are the
resonance dispersions calculated  using the parameters listed in
the text.}\label{fig7}
\end{figure}

	Finally, we present the results of magnetotransmission
measurements done on sample (II) at the High Magnetic Field
Laboratory in Grenoble . In Fig.~\ref{fig7} we display the
magnetic-field dispersion of resonances between 15 and 28 T for
unpolarized radiation.  The lines are the calculated dispersions
using the same parameters as above, but now calculated up to B=30
T.  The most important result from these additional high magnetic
field experiments is that they have permitted us to experimentally
observe the polaron transition at energies above the restrahlen
band of the substrate.  Indeed at 20 T, we start to see an
absorption at $\sim$40 meV. This is a significant result because
this energy transition cannot be predicted using a purely
electronic model. We therefore have additional experimental
support that the energy transitions occur between polaron states.
Note that in Fig.~\ref{fig7} the calculated upper branch is found
to be $\sim\text{2 meV}$ below the experimental points.  Such a
discrepancy could, e.g., arise from an interaction involving the
LO-phonon of GaAs ($\sim\text{36 meV}$) which is not included in
our calculations where only the InAs like LO-phonon is taken into
account.

\section{Conclusion}
In summary, we have investigated the valence intraband transitions
in p-doped self-assembled InAs QDs by using FIR magneto-optical
technique with linearly polarized radiation. We have shown that a
purely electronic model is unable to account for the experimental
data, neither for the energy dispersion of the magneto absorption
resonance, nor for the intensity dependence versus magnetic field.
As the transition energies are close to that of the LO-phonon in
InAs, we have shown that a model taking into account the hole
LO-phonon coupling is able to predict well the experimental data.
We have calculated the coupling between the relevant mixed
hole-lattice states using the Fr\"ohlich Hamiltonian and we have
determined the polaron states as well the oscillator strength of
the polaron transitions as a function of magnetic field.  We
believe that the fact that our model successfully fits the
experimental data constitutes the first evidence for the existence
of hole polarons in InAs QDs and demonstrates that the intraband
magneto-optical transitions occur between hole polaron states.
 \acknowledgments
The Laboratoire Pierre Aigrain is a mixed Research Unit (UMR 8551)
between Ecole Normale Superieure, the University Pierre et Marie
Curie (Paris6) and the CNRS. We would like to thank G. Bastard for
very valuable and fruitful discussions.
\appendix

\section{Calculating polaron levels}
In order to understand the experimental results observed in Fig.
3, we develop a simple model that involves dispersionless LO
phonons.  We deal with the zero-phonon discrete hole levels
$|p+,0\rangle$, and $|p-,0\rangle$ and the flat one-phonon
continuum $|s,1\rangle$, as well as the states $|s,0\rangle$,
$|s,2\rangle$ and $|p\pm,1\rangle$ which are treated solely in
perturbation. The Fr\"ohlich term of the Hamiltonian is written as
follows,

\begin{equation}\label{Frohlich_term}
  H_{int}=\sum_{\textbf{q}}[{V(\textbf{q})a^{+}_{q}+V^{*}(\textbf{q})a_{\textbf{q}}}],
\end{equation}
where $\textbf{q}$ are the different modes of the LO-phonon and
$V(\textbf{q})$ includes in its definition the dimensionless
Fr\"ohlich constant $\alpha$. First, we consider the interaction
between one discrete level, say $|p-,0\rangle$, and the one phonon
continuum $|s,1\rangle$. We note $V_{sp-}(\textbf{q})=\langle
s,1_{\textbf{q}}| H_{int}|p-,0\rangle$. We perform linear
combinations of the degenerate one-phonon states
$|1_{\textbf{q}}\rangle$ and introduce a particular linear
combination,
\begin{equation}\label{linearcombo1}
  |1_{\alpha(sp-)}\rangle=\frac{\sum_{\textbf{q}}V_{sp-}(\textbf{q})|1_{\textbf{q}}\rangle}{\sqrt{\sum_{\textbf{q}}|V_{sp-}(\textbf{q})|^{2}}},
\end{equation}
\begin{equation}\label{linearcombo2}
\langle s,1_{\alpha(sp-)}|H_{int}|p-,0\rangle =
  \sqrt{\sum_{\textbf{q}}|V_{sp-}(\textbf{q})|^{2}}=\upsilon_{eff}(sp-).
\end{equation}
We label $|1_{\beta}\rangle$ the remaining one-phonon combinations
orthogonal to $|1_{\alpha(sp-)}\rangle$.  Thus, the continuum
levels $|s,1_{\beta}\rangle$ are uncoupled to $|p-,0\rangle$ .  We
have in this way an interaction between two discrete levels that
gives rise to two polaron states.  These polarons are a mixture of
$|p-,0\rangle$ and $|s,1_{\alpha(sp-)}\rangle$.  We can now
generalize this procedure to handle the interactions between
several mutually orthogonal discrete zero phonon levels and a flat
continuum with one LO phonon: each discrete level is coupled to
only one particular linear combination of each continuum.  So if
we have $N$ discrete levels and $M$ continuums, we need only to
consider $N\times(M+1)$ levels.  In the present case, we have
$N$=2 ($|p+,0\rangle$ and $|p-,0\rangle$) and $M$=1
($|s,1\rangle$). We therefore expect to find four polaron levels.
In this basis, the polaron wavefunction is written:

\begin{multline}\label{wavefunction}
|\Psi\rangle=C_{p+}|p+,0\rangle+C_{p-}|p-,0\rangle \\
 +C_{sp+}|s,1_{\alpha(sp+)}\rangle+C_{sp-}|s,1_{\alpha(sp-)}\rangle
\end{multline}
where $|1_{\alpha(sp-)}\rangle$ is given in
eqn.~\ref{linearcombo1} and $|1_{\alpha(sp+)}\rangle$ is obtained
in a similar way.  The corresponding eigenvalue equation is

\begin{equation}\label{eigenvalue_equation}
\begin{pmatrix}
\tilde{\varepsilon}_{s1}-E&\upsilon_{eff}&0&0\\
\upsilon_{eff}&\tilde{\varepsilon}_{p_{+}0}-E&0&\delta_{p}\\
0&0&\tilde{\varepsilon}_{s1}-E&\upsilon_{eff}\\
0&\delta_{p}&\upsilon_{eff}&\tilde{\varepsilon}_{p_{-}0}-E
\end{pmatrix}
\begin{pmatrix}
C_{sp+}\\C_{p+}\\C_{sp-}\\C_{p-}
\end{pmatrix}=\overrightarrow{0}
\end{equation}
\\
where $\delta_{p}$ is the anisotropy term that mixes the 0-phonon
states of $p$ symmetry and

\begin{equation}\label{s_energy}
\tilde{\varepsilon}_{s1}=E_{s}(B)+\hbar\omega_{LO}-\frac{\sum_{\textbf{q}}|V_{ss}(\textbf{q})|^{2}}{\hbar\omega_{LO}},
\end{equation}

\begin{equation}\label{p_energy}
\tilde{\varepsilon}_{p_{\pm}0}=E_{p_{\pm}0}(B)-\frac{\sum_{\textbf{q}}|V_{pp}(\textbf{q})|^{2}}{\hbar\omega_{LO}},
\end{equation}

\begin{equation}\label{v_eff}
\upsilon_{eff}=\sqrt{\sum_{\textbf{q}}|V_{sp}(\textbf{q})|^{2}},
\end{equation}

\begin{equation}\label{perturbation_term}
V_{ij}(\textbf{q})=\langle i|V(\textbf{q})|j\rangle.\\
\end{equation}
\\
$E_{j}(B)$ is the $B$-dependent energy of the 0-phonon electronic
level $|j\rangle$, which contains both a diamagnetic $(\sim
B^{2})$ and Zeeman $(\sim B)$ term.  The last term in
$\tilde{\varepsilon}$ is a second order perturbation correction
resulting from the interaction of $|s,0\rangle$ and $|s,2\rangle$
with $|s,1\rangle$ and $|p\pm,0\rangle$ with $|p\pm,1\rangle$.

	Note that a dot with cylindrical symmetry $(\delta_{p}=0)$ has
two independent sets of polaron levels, whose field dispersions
are obtained by solving
\begin{equation}\label{no_anisotropy1}
(\tilde{\varepsilon}_{s1}-E)(\tilde{\varepsilon}_{p_{+}0}-E)=\upsilon_{eff}^{2}
\end{equation}
or
\begin{equation}\label{no_anisotropy}
(\tilde{\varepsilon}_{s1}-E)(\tilde{\varepsilon}_{p_{-}0}-E)=\upsilon_{eff}^{2}.
\end{equation}

	Also note that in the absence of Fr\"ohlich coupling
($V_{ij}=0$), eqn.~\ref{eigenvalue_equation} leads to two
dispersion curves for the $|p\pm,0\rangle$ states, solutions of
\begin{equation}\label{no_frohlich}
[E_{p_{+}0}(B)-E][E_{p_{-}0}(B)-E]=|\delta_{p}|^{2}
\end{equation}
which we see in Fig.~\ref{fig4a} in dashed lines.

	In the general case, the discrete states are at the same time
coupled to each other and to the continuum states. However, since
the anisotropy term is weak in our samples
($\delta_{p}\simeq0.9$meV), the resulting polarons are roughly
described in terms of the coupling of $|s,1_{\alpha(sp-)}\rangle$
with $|p-,0\rangle$ and $|s,1_{\alpha(sp+)}\rangle$ with
$|p+,0\rangle$.

	The optical absorptions detected in our measurements (see
Fig.~\ref{fig2}) involve the excitation of the hole in the quantum
dot from the ground state $|s,0\rangle$ towards the set of four
polaron levels $|\Psi\rangle$ resulting from the diagonalization
of the $4\times4$ matrix in eqn.~\ref{eigenvalue_equation}.  For
light polarized in the QD layer plane the oscillator strength for
each transition is proportional to
\begin{equation}\label{OS}
OS_{|\Psi\rangle}=|(C_{p+}+C_{p-})\epsilon_{[110]}+(C_{p+}-C_{p-})\epsilon_{[1\overline{1}0]}|^{2},
\end{equation}
where $\epsilon$ is the in-plane polarization direction of the
electromagnetic wave. We show in Fig.~\ref{fig6} the field
dependence of the oscillator strength for the two low-lying
polaron levels for both [110] and [1$\overline{1}$0]
polarizations.



\newpage




\begin{thebibliography}{99}


\bibitem{hameau02}
S. Hameau, J.N. Isaia, Y. Guldner, E. Deleporte, O. Verzelen, R.
Ferreira, G. Bastard, J. Zeman, and J.M. G\'erard, Phys. Rev. B
{\bf 65}, 085316 (2002).

\bibitem{sarkar05}
D. Sarkar, H.P. van der Meulen,J.M. Calleja, J.M. Becker, R.J.
Haug, and K. Pierz Phys. Rev. B {\bf 71},081302(R) (2005).

\bibitem{knipp97}
P.A. Knipp, T.L. Reinecke, A. Lorke, M. Fricke, and P.M. Petroff,
Phys. Rev. B {\bf 56}, 1516 (1997).

\bibitem{hameau99}
S. Hameau, Y. Guldner, O. Verzelen, R. Ferreira, G. Bastard, J.
Zeman, A. Lema\^\i tre, and J.M. G\'erard, Phys. Rev. Lett. {\bf
83}, 4152 (1999).

\bibitem{verzelen00}
O. Verzelen, R. Ferreira, and G. Bastard, Phys. Rev. B {\bf 62},
R4809 (2000).

\bibitem{inoshita97}
T. Inoshita and H. Sakaki, Phys. Rev. B {\bf 56}, R4355 (1997).

\bibitem{li98}
X.Q. Li and Y. Arakawa, Phys. Rev. B {\bf 57}, 12285 (1998).

\bibitem{Goldstein85}
L.Goldstein, F. Glas, J.Y. Marzin, M.N. Charasse, and G. Leroux,
Appl. Phys. Lett. {\bf 47}, 1099 (1985).

\bibitem{fricke96}
M. Fricke, A. Lorke, J.P. Kotthaus, G. Medeiros-Ribeiro, and P.M.
Petroff, Europhys. Lett. {\bf 36}, 197 (1996).

\bibitem{hasegawa98}
Y. Hasegawa, H. Kiyama, Q.K. Xue, and T. Sakurai, Appl. Phys.
Lett. {\bf 72}, 2265 (1998).

\bibitem{stier01}
O. Stier,\textit{ Electronic and Optical Properties of Quantum
Dots and Wires} (Wissenschaft und Technik Verlag).

\bibitem{bester03}
G. Bester, S. Nair, and A. Zunger, Phys. Rev. B {\bf 67},161306(R)
(2003).

\bibitem{ferreira99}
R. Ferreira, and G. Bastard, Appl. Phys. Lett. {\bf 74}, 2818
(1999).




\bibitem{Heitz97}
R. Heitz, M. Veit, N.N. Ledentsov, A. Hoffmann, D. Bimberg, V.M.
Ustinov, P.S. Kop'ev, and Zh.I. Alferov, Phys. Rev. B {\bf 56}, 10
435 (1997).











\end{thebibliography}
\end{document}